\documentclass[12pt,preprint]{aastex}
\shorttitle{Secular evolution in Mira pulsations}
\shortauthors{Templeton et al.}

\begin{document}
\title{Secular Evolution in Mira Variable Pulsations}
\author{M. R. Templeton and J. A. Mattei}
\affil{American Association of Variable Star Observers, Clinton B. Ford
	Astronomical Data and Research Center, 25 Birch Street, Cambridge, 
	MA 02138}
\author{L.A. Willson}
\affil{Department of Physics and Astronomy, Iowa State University, Ames, IA
	50011-3160}

\begin{abstract}
Stellar evolution theory predicts that asymptotic giant branch stars
undergo a series of short thermal pulses that significantly change their
luminosity and mass on timescales of hundreds to thousands of years.  
These pulses are confirmed observationally by the existence of the 
short-lived radioisotope technetium in the spectra of some of these stars, but
other observational consequences of thermal pulses are subtle, and may only
be detected over many years of observations.  Secular changes in these stars 
resulting from thermal pulses can be detected as measurable changes in period
if the star is undergoing Mira pulsations.  It is known that a small fraction
of Mira variables exhibit large secular period changes, and the detection of 
these changes among a larger sample of stars could therefore be useful in 
evolutionary studies of these stars.  The {\em American Association of Variable
Star Observers} (AAVSO) International Database currently contains visual 
data for over 1500 Mira variables.  Light curves for these stars span nearly a
century in some cases, making it possible to study the secular evolution of 
the pulsation behavior on these timescales.  In this paper, we present the
results of our study of period change in 547 Mira variables using data from the 
AAVSO.  We use {\it wavelet analysis} to measure the period changes in 
individual Mira stars over the span of available data.
By making linear fits to the period versus time measurements, we 
determined the average rates of period change, $d\ln{P}/dt$, for each of these
stars.  We find non-zero $d\ln{P}/dt$ at the 2-$\sigma$ significance level
in 57 of the 547 stars, at the 3-$\sigma$ level in 21 stars, and at the level of
6-$\sigma$ or greater in eight of the 547.  The latter eight stars have been
previously noted in the literature, and our derived rates of period changes 
largely agree with published values.  The largest and most statistically
significant $d\ln{P}/dt$ are consistent with the rates of period change 
expected during thermal pulses on the AGB.  A number of other stars exhibit
non-monotonic period change on decades-long timescales, the cause of which is 
not yet known.  In the majority of stars, the period variations are smaller 
than our detection threshold, meaning the available data are not sufficient 
to unambiguously measure slow evolutionary changes in the pulsation period.
It is unlikely that more stars with large period changes will be found among
heretofore well-observed Mira stars {\it in the short term}, but continued
monitoring of these and other Mira stars may reveal new and
serendipitous candidates in the future.
\end{abstract}


\section{Introduction}

The asymptotic giant branch (AGB) is an important evolutionary stage for
stars of low to intermediate mass.  It represents the most luminous stage
of these stars' lives, and the stage during which they begin to return a 
significant fraction of their mass to the interstellar medium \citep{law00}.
The AGB is also
the location of the Mira variables, which are high-amplitude, pulsating 
variable stars; as with several other types of pulsating variables, the Miras 
are also useful as distance indicators \citep{fwm02}.  Thus observational 
studies of stars on the AGB, particularly the pulsating stars, provide 
information useful for several other important astrophysical 
problems.

The Mira stars are also reasonably easy targets of study.  They make up a 
significant percentage of the catalogued variable stars in the nearby universe
\citep{gcvs}.  Their large luminosities 
and large amplitudes make them easy to detect and monitor, and their long 
periods allow for easy study since observations can be widely spaced 
in time.  Many hundreds of Mira variables have been regularly observed by
amateur and professional astronomers for over a century, and the long spans of
data collected on these stars now enable the study of {\it secular changes} in 
pulsation behavior.  These secular changes can potentially provide useful
information on several aspects of Mira variable and AGB evolution, including 
thermal pulses and pulsation-driven mass-loss.

Mira variables have long been known to have varying periods and other aperiodic
behavior.  Plakidis and Eddington (\citet{ep29,p32}) investigated the effects 
of both measurement error and intrinsic cycle-to-cycle variations on the $O-C$
diagrams of Mira stars and LPVs, and \citet{p32} states that ``...the periods 
of [R Aql and R Hya] are well-known to decrease.''  R Aql and R Hya were again
identified as stars with statistically significant period changes by 
\citet{sc37}.  The ability to detect period changes increased along with the 
span of available observations, and by the 1970's, very large data sets
could be used to detect and study such behavior.  Since the 1980's, work
in this field has blossomed; for example, \citet{wz81} noted that W Dra was 
a candidate for period changes, \citet{aa85} noted a significant period 
increase in LX Cyg, and \citet{b88}
found the same in BH Cru.  The most spectacular case of period change in
a Mira star is that of T UMi, described independently by \citet{gs95} and
\citet{mf95}.  Since then, several surveys and discussions of period
change have appeared in the literature, and it is now generally 
(but not universally) accepted that large period changes are caused by 
{\it thermal pulses}.

The existence of thermal pulses follows from standard stellar evolution 
models \citep{wz81,ir83,bs88,vw93}, 
and also from the observation of technetium in the spectra of some AGB stars 
\citep{lh99}.  However, the consequences of thermal pulses should 
also be observable in the {\it pulsation behavior} of variable stars on the 
AGB.  The expected rate of period change of the fundamental mode is given by

\begin{equation}
{d\ln{P} \over dt} = -0.32 {d\ln{M} \over dt} + 0.55 {d\ln{L} \over dt}
\end{equation}

\noindent
where $d\ln{P}$, $d\ln{M}$, and $d\ln{L}$ are the period, mass, and luminosity
normalized by the average value over the interval $dt$.  This equation is 
obtained by combining the fundamental mode period-mass-radius relation of 
\citet{oc86} with the relation $d\log{R} / d\log{L} = 0.68$ from 
\citet{iben84}.  A similar equation results for overtone pulsations or for 
constant $Q = P \sqrt{\rho/\rho_{\odot}}$, but with somewhat smaller 
coefficients.  The \citet{wz81} interpretation of R Aql and R Hya as shell 
flashing giants assumed overtone pulsation, but recent observations make it 
clear that Miras are fundamental mode pulsators \citep{p2004,w2004}.  For the 
Mira variables, $d\ln{L}/dt \sim -d\ln{M}/dt$, so both mass loss and luminosity
changes will significantly affect pulsation periods on the AGB.  During thermal
pulses, the rates of mass and luminosity change may increase on secular 
timescales, causing detectably large changes in the period.

First, thermal pulses will change the stellar luminosity and temperature, and
readjust the interior structure of the star on the 
thermal timescale.  Evolution models predict that for a short period of 
time after the pulse onset (less than 1000 years), large and 
observable changes will occur to the star.  Such changes would
manifest themselves as changes in pulsation behavior, because of changes 
in both the interior sound speed and stellar radius.
Mass loss also occurs on the AGB, and may affect pulsations in Mira variables
on observable timescales.  First it would cause a decrease in envelope 
mass, which affects the pulsation period(s) as shown in equation (1).  
Second, it may also cause an increase in dust and molecular opacity, which 
would affect the absolute magnitude of the star and change the spectral energy
distribution.  Third, \citet{iben84} evolutionary models (for example) predict
that as the mass decreases, the radius {\it increases} (as per equation 1 
above) during late states of evolution for stars more massive than about 1.16 
M$_{\odot}$.  The mass loss rates of these stars can be as high as a few 
times 10$^{-5}$ M$_{\odot}$/yr, affecting the pulsation period over the course
of a century even though the amount of mass lost would be relatively small (a 
few thousandths of a solar mass).  

According to most evolutionary models \citep{wz81,bs88,vw93}, the largest 
period changes caused by thermal pulses should occur over relatively short 
periods of time, perhaps a few
thousand years at most.  The interpulse phases of AGB evolution may last for
a few hundred thousand years.  If we assume that Mira pulsations can occur 
during most of the pulse {\it and} interpulse phases, then we
might expect that a few percent of all observed Mira variables exhibit
significant changes to their pulsation behavior over the span of available
observations.  These changes may manifest themselves as changes to (i) 
pulsation period, (ii) mean magnitude, and (iii) amplitude and/or light curve
shape.

A third possible reason for secular changes in Mira variable pulsations 
unrelated to evolutionary processes is nonlinear behavior.  In most other 
types of 
pulsating variable, the pulsations have small enough amplitudes that they do 
not significantly deviate from strictly periodic behavior, and do not impact 
the equilibrium structure of the star.  This is why linear perturbation
theory has been so effective in modeling stellar pulsations in general.
However, for the Mira stars this is not necessarily the case.  It is well-known
that the Miras have cycle-to-cycle variations in period and amplitude, 
suggesting nonlinear behavior.  Nonlinearity can also affect the propagation
of atmospheric shocks into regions forbidden by linear theory, where
$P > P_{\rm acoustic}$ \citep{b90}, leaving open the possibility of additional
nonlinear effects. Some work has also been done to determine
whether chaos plays a role in Mira pulsations \citep{can90,bed98,ks02},
although not all of these studies confirmed chaotic behavior.  The semi-regular 
and long-period irregular variables are in similar evolutionary stages as the
Miras, and nonlinear dynamics and chaos may be at work in these stars as well.
Any studies of Mira variable evolution based on pulsation behavior must keep 
this in mind.  More modeling work along these lines is clearly warranted.  

In this paper, we present a time-series analysis of 547 Mira variable star 
light curves from the {\em American Association of Variable Star Observers}
(AAVSO) 
International Database (ID), to search for evolutionary secular changes among 
Mira stars.  We use the technique of {\em wavelet analysis} to measure changes
in the pulsation periods of these stars over the span of available 
observations.  We then investigate the stars which exhibit statistically 
significant changes in period to assess whether
these changes may have arisen from evolutionary events.  In section 2 of this
paper, we describe the AAVSO data and our sample selection criteria.  In
section 3, we describe our wavelet analysis algorithm and procedure for finding
statistically significant period variations.  In section 4, we describe the
results of our time-series analysis and compare our results with those 
previously published.  In section 5, we discuss
the implications of our results on our understanding of Mira stars and AGB
evolution.

\section{Data}

In this work, we make use of visual magnitude estimates collected and archived
by the AAVSO.  These
visual magnitude estimates were made primarily by amateur astronomers using
small telescopes or binoculars.  Because individual stars may have hundreds 
of different observers contributing observations, the data are necessarily 
heterogeneous due to variations in the response of the eyes of different 
individuals and in the type of observing aid used.  However, the amount of
data available for many of these stars is so large that the resulting averaged
light curves are of very high quality.  Furthermore, all observers used
the same comparison star sequence, which improves the consistency of data
taken by different observers.  The statistical properties of the data
within a given light curve are largely constant over time, so the data
may be treated as statistically homogeneous.  The data have also undergone 
a validation process by which clearly discrepant points are flagged, checked 
by AAVSO staff, and removed from the light curves (if need be) prior to 
analysis \citep{wm03}.  Finally, the visual data are averaged into 10-day bins
to produce the final light curve which we then analyze.  

For this paper, we selected stars for analysis based upon two criteria: the 
AAVSO ID must contain at least 500 valid data points for the star, and the 
most recent version of the {\em General Catalogue of Variable Stars}
\citep{gcvs} must classify the object as Mira (M) or suspected Mira (M:). Using
these criteria, the data for 547 stars were selected from the AAVSO ID.  

\begin{figure}
\figurenum{1}
\label{fig1}
\epsscale{0.5}
\plotone{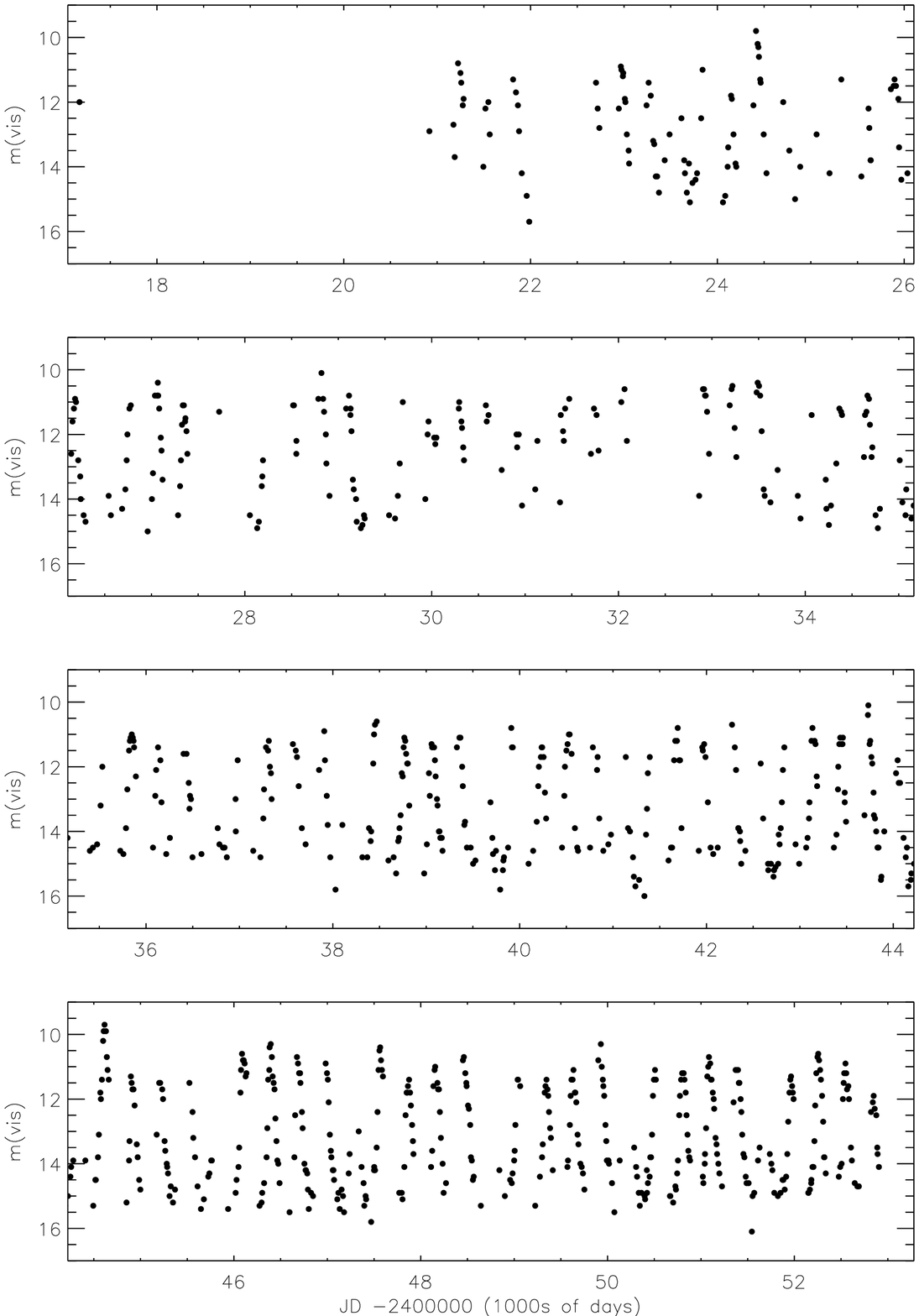}
\caption{Light curve of 10-day means of S Arietis 
(P = 292 days).  The
light curve was obtained from visual observations, and spans nearly 90 years.
This light curve highlights some of the features and drawbacks of the AAVSO
data.  The light curve coverage is sparse up to $\sim JD 2435000$, but
markedly improves later on, as an increasing number of observers contribute 
their observations.  After $JD 2440000$, the maxima and minima are
well-covered, showing the modest variability in the magnitudes of maximum and
minimum light.}
\end{figure}

\begin{figure}
\figurenum{2}
\label{fig2}
\epsscale{0.75}
\plotone{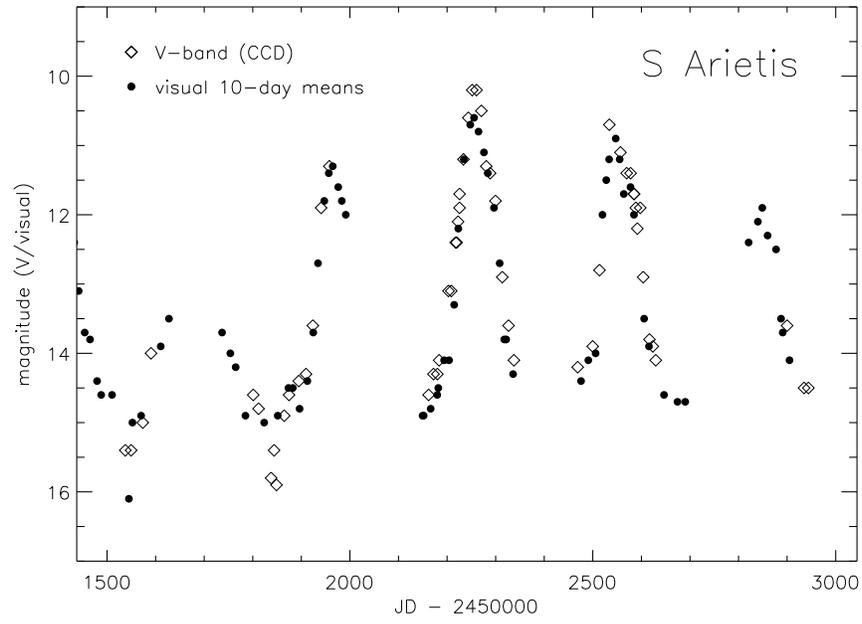}
\caption{10-day means of S Ari versus $V$-band CCD observations.  The agreement
between the CCD and visual data is good, indicating that the visual data
are reliable estimators of the star's photometric behavior.}
\end{figure}

As an example, Figure \ref{fig1} shows the 10-day mean light curve of visual 
observations of S Arietis, an ordinary Mira variable with moderate light curve
coverage typical of our sample.  The light curve shows
both the strengths and weaknesses of our data.  The light curves track the 
behavior of this variable well during most of the past 90 years, particularly 
in the coverage of maxima and minima which show amplitude variations.  
However, increasingly poor coverage and larger data gaps earlier in the data 
set make it more difficult to track the earliest behavior of this object.  
Furthermore, even the most recent data are not so well-covered that the 
variations in times and magnitudes of maxima are exactly defined, and there is 
still scatter of a few tenths of a magnitude in the brightnesses of minima.  
Figure \ref{fig2} shows a portion of the visual light curve overlaid with
CCD $V$-band measurements of this star, also taken by AAVSO observers.
Although the two data sets agree for a substantial portion of the light curves,
there are significant differences in places, notably near JD 2451850 (minimum)
and JD 2452250 (maximum).  Such differences were noted by \citet{z03}, who 
compared AAVSO visual observations to calibrated $B$-,$V$-, and $R$-band data 
taken specifically
for transformation purposes, and noted that several factors including color
differences between comparison stars and physiological differences between
observers can contribute.  He also noted that such effects can make 
transformation between visual magnitude estimates and standard photometric
systems very difficult.  However, we are most concerned that the visual 
observations represent the general behavior of the star properly, regardless
of whether their absolute calibration is correct, and the visual and
CCD data are in clear qualitative agreement in the behavior of the star; the 
same conclusion was reached by \citet{lk01}.  Therefore the visual observations
form a valid data set for performing this kind of study.

\begin{figure}
\figurenum{3}
\label{fig3}
\epsscale{0.75}
\plotone{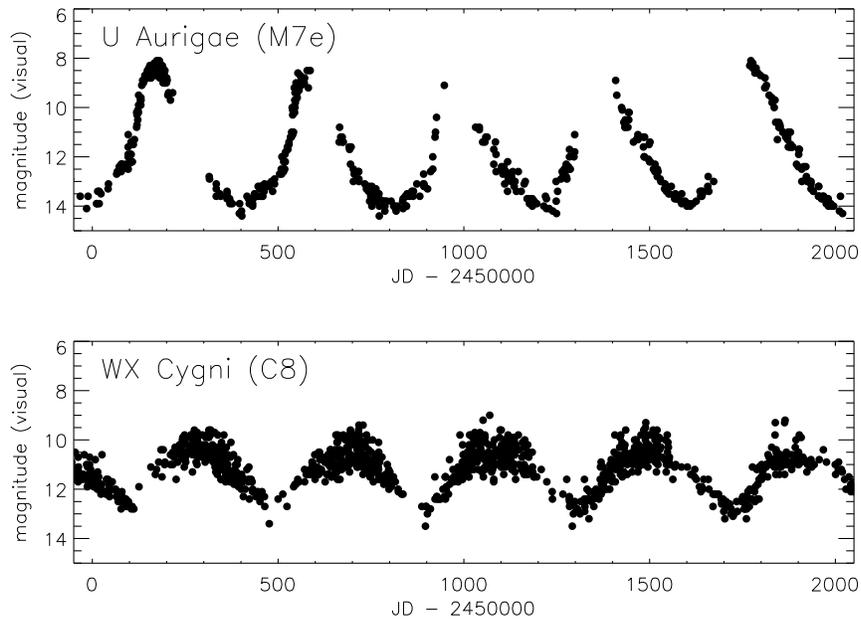}
\caption{Raw visual light curves of U Aur and WX Cyg, M- and C-type Mira 
variables, respectively.  The larger scatter in visual observations of WX Cyg 
is due to this star being redder (B-V = 3.2; \citet{alk01}) than U Aur 
(B-V = 2.1).  Light curves of very red stars are more susceptible to 
observational effects such as the Purkinje effect.}
\end{figure}

The intrinsic 1-$\sigma$ errors of visual brightness estimates are 
approximately 
0.1 to 0.3 magnitudes, with larger scatter commonly observed for redder stars.
Because the data are not evenly sampled, binning of data may result in 
varying RMS errors per point along the light curve, but the {\it average} RMS 
error per point is between 0.1 and 0.3 magnitudes.
Figure \ref{fig3} shows the raw light curves for two Mira variables having approximately
the same period, but very different spectral types.  The carbon Mira WX Cygni
shows significantly more scatter in its light curve than does the oxygen Mira 
U Aurigae.  This increased scatter arises from a combination of factors, 
including differences in ocular red-sensitivity between observers, observations
with equipment of different aperture sizes, different observing techniques, and
ultimately, the Purkinje effect.  For a longer discussion of this topic, we
refer the reader to \citet{h98} and \citet{s99}.  Because of the 
increased scatter observed in redder stars, we expect higher uncertainties in 
the data for S- and C-type Miras, particularly those with lower amplitudes.

\section{Analysis method}

To search for period changes, we use a version of the
{\em weighted wavelet Z-transform} ({\em WWZ}) developed by \citet{f96} 
specifically for analyzing AAVSO data.  The {\em WWZ} algorithm projects the 
data onto a set of sine and cosine trial functions (the wave form); 
a test sine wave having fixed frequency is fit to the data using 
the Gaussian wavelet window function 

\begin{equation}
W(\omega,\tau) = exp(-c \omega^{2} (t - \tau)^{2})
\end{equation}

\noindent
as the weighting function of the data, where $\omega$ is the test frequency,
$\tau$ is the center of the test window, and $c$ is a tuning constant used to
adjust the width of the window.  The algorithm was
designed with irregularly sampled data in mind, and is analogous to 
date-compensated Fourier methods such as the \citet{fm81}
date-compensated DFT, and the Lomb-Scargle periodogram \citep{s82}.
\citet{hmf01} studied the secular evolution of R Cen using the 
{\em WWZ} algorithm, and they showed that {\em WWZ} produces similar
results to dividing the data into finite segments and performing a
date-compensated DFT on each segment.  {\em WWZ} is preferable to manually
dividing the data, as the former automatically selects how much data to include
in each iteration based upon the test frequency and the window width, rather
than having a fixed span of data at a given time step.

For our tests, we step the wavelet window over the data at time intervals, 
$\Delta \tau$, 
of 500 days, and scan a frequency range of 0.0005 to 0.02 cycles per day 
(corresponding to periods of 2000 to 50 days respectively).  The Gaussian 
window width is 
frequency-dependent, such that the number of cycles of the test frequencies 
appearing within the window is constant.  We used a tuning constant,
$c$, of 0.001, to smooth out the cycle-to-cycle variations and emphasize the 
longer-term trends in pulsation behavior.  A value of 0.001 sets the e-folding
full-width of the weighting window at approximately 20 cycles; at a distance of
ten cycles on either side of the wavelet window center, the data weights have 
fallen from $1$ to $1/e$.  We tested several values of $c$ between 0.012 and 
0.0001, and found 0.001 gave the best results for our purposes.  Values smaller
than 0.001 tended to wash out all of the variations including the long-term 
ones because smaller values of $c$ result in larger windows.  Increasing the 
value of $c$ decreased the frequency resolution and increased the scatter in 
peak frequency, making it difficult to track subtle changes.

As \citet{f96} noted, {\em WWZ} is well-suited to detecting
period variations but not amplitude or mean magnitude changes.  This problem
is most acute where the data contain gaps due to solar interference, seasonal
weather variations, or lack of observers.
For example, the amplitude and mean magnitude are often 
underestimated due to missed minima; many of the Mira variables in our sample
have minima below $m_{\rm vis} = 15$, beyond the capabilities of observers 
with small- to moderate-sized telescopes.  {\em WWZ} also 
underestimates total amplitude even for well-sampled, regular light curves 
because our particular implementation of the transform assumes purely 
sinusoidal variations.  {\em WWZ} detects higher-order Fourier harmonics,
but our implementation does not compute the phase information and so the 
Fourier harmonics were not combined to determine the total light amplitude.

Following the automated wavelet analysis of the 547 light curves, we inspected
the results of each by eye.  For our analysis, we require only the peak
period as a function of lag time, $\tau$, rather than the full spectrum.  
We found that 89 of the 547 stars showed either noise spikes or sharp 
discontinuities in plots of peak period versus time.  In most cases,
these points are spurious, due to gaps in the data.  In a few
cases (including DH Cyg, DU Cyg, R Nor, and R Cen), the light curves are or 
appear to be double-peaked, and the discontinuities result from {\em WWZ}
selecting the shorter period of the smaller peaks rather than the longer
period of the main variation.  We removed these points before we 
analyzed the {\em WWZ} output for period changes.

Finally, we estimated the average rate of period change for all the stars in
the sample by fitting a straight line through the values of $P(\tau)$, and
obtaining $d\ln{P}/dt$ from the slope of the line.  In some cases, assuming a 
linear period change is clearly incorrect.  In the unique case of T UMi
(discussed below), we fit a line through the epoch of declining period only.
For those stars that do not 
exhibit monotonic period change (such as RU Tau and S Ori) we also
compute the largest deviations from the mean period during the span of the 
observations.  For both calculations, we assume that the error in period is
the half-width at half-maximum of the {\em Z-statistic}, $Z(\omega,\tau)$
\citep{f96}.  We measured this for several test stars, and found that the 
half-width at half-maximum
was typically on the order of 1-2\% of the period.  Therefore, we 
conservatively assumed that the 1-$\sigma$ error bars, $\sigma_{P}$, of 
individual measurements of $P(\tau)$ at each timestep are fixed at 
$0.02 P(\tau)$ for all stars at all times.  In reality, $\sigma_{P}$ varies 
as a function of the number of data points within a given window because the 
coverage of the individual light curves is not uniform.  However, our use of 
an average error for simplicity is reasonable; we exclude regions of the light
curves where the data are very poorly-covered, and the value of $0.02 P(\tau)$
is likely overly conservative for well-covered light curves.  Therefore,
we are confident
we are not drastically underestimating the errors in period, but may be 
overestimating them for well-observed stars.

Once the curves of $P(\tau)$ versus $\tau$ were obtained, we made linear
fits to each star's curve and obtained a slope, $dP/dt$, standard
deviation of the slope, $\sigma_{\rm fit}$, and average period, $\bar{P}$.
The rate of period change, $d\ln{P}/dt$, is then defined as the slope divided
by the average period, and the significance level was obtained by dividing
the slope by $\sigma_{\rm fit}$.  Stars with significance levels greater
than two were flagged, and those greater than three were considered 
significant.

\section{Results}

\begin{figure}
\figurenum{4}
\label{fig4}
\epsscale{0.75}
\plotone{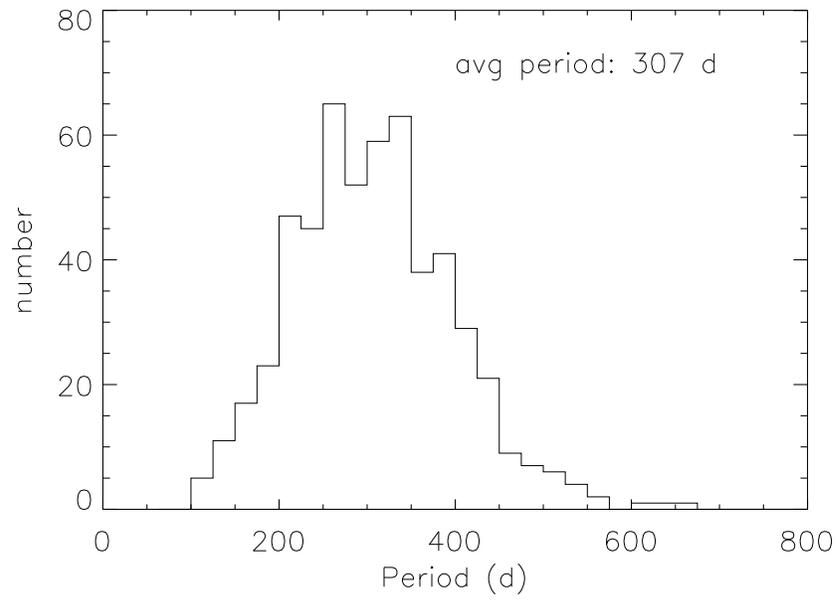}
\caption{Histogram of the 547 Mira variables in our sample, binned by period.
The average period of our sample is 307 days.}
\end{figure}

\begin{figure}
\figurenum{5}
\label{fig5}
\epsscale{0.75}
\plotone{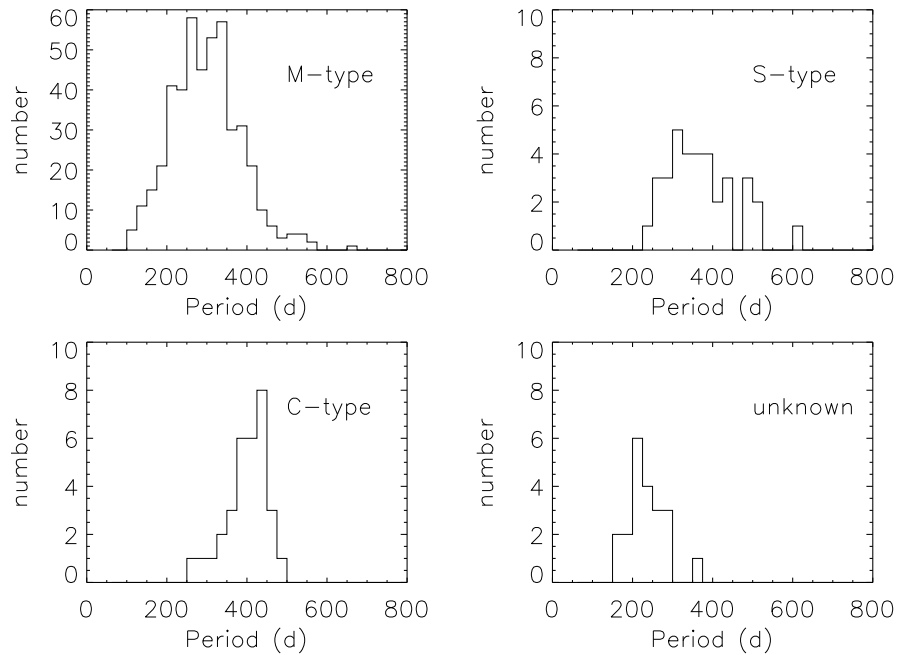}
\caption{Histograms of the period distribution of the 547 Miras in our sample,
subdivided by spectral type.  The M- and S-type Miras have broad distributions
of period, while the C-type Miras are more densely concentrated at longer
periods.  When the period distributions are compared to those of all Miras 
listed in the {\it GCVS}, the AAVSO sample of C-type Miras appears to be 
biased relative to the period distribution of the {\it GCVS} C-type sample, 
but the M- and S-types in our sample are not statistically different from the 
{\it GCVS} sample.}
\end{figure}

Our main goal is to search for period changes in Mira stars and determine
whether they are caused by evolutionary changes within the star.  First,
a preliminary discussion of the 
global properties of the sample is in order.  Figure \ref{fig4} shows the 
number distribution of periods within our sample.  Overall, the sample has a 
wide distribution in period, with an average period of 307 days.  Figure 
\ref{fig5} shows the same histogram divided into the three subtypes (M-, S-, 
and C-type Miras) and those that are unclassified.  The majority of the Miras
in our sample are M-type (458), with the S-type (35), C-type (32), and 
unclassified Miras (21) making up a much smaller fraction.   The S- and C-type
Miras also have longer average periods than the M-types; the S- and C-types 
have average periods of 373 and 400 days, respectively, while the M-types have
an average period of 298 days.  The period distributions of the M- and S-type
Miras in our sample agree with those of all Miras in the 
{\em GCVS}, both in terms of mean period and of the distribution of short-,
medium-, and long-period Miras of each sub-type.  However, our sample of
C-type Miras is underabundant in short-period stars relative to the {\em GCVS}
distribution.

\begin{figure}
\figurenum{6}
\label{fig6}
\epsscale{1.00}
\plotone{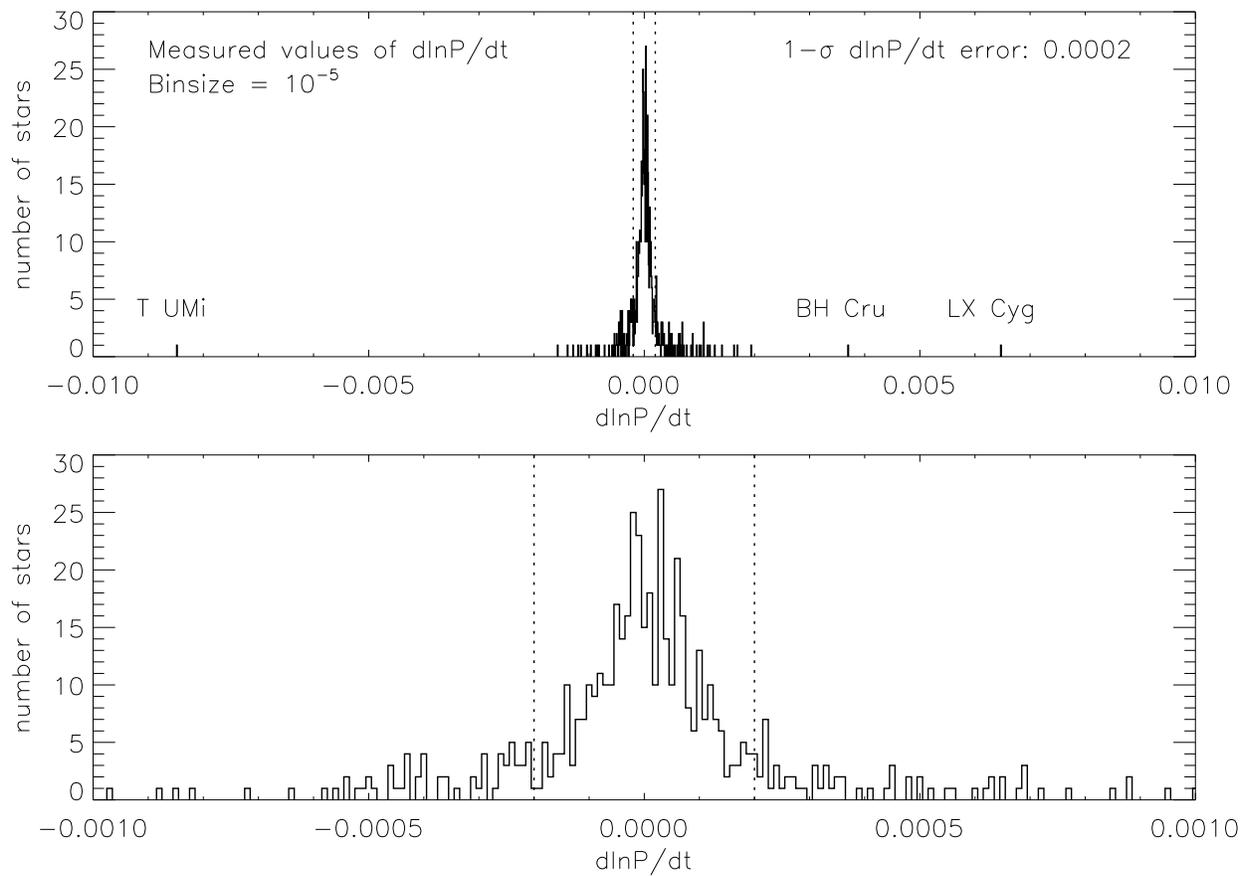}
\caption{Histograms of $(d\ln{P}/dt)$ for the 547 stars in our sample.  The
1-$\sigma$ error in $(d\ln{P}/dt)$ is marked by the vertical dashed lines.
The majority of stars in our sample have $(\vert{d\ln{P}/dt}\vert)$ less than
this, and thus do not have measurable period changes given the current data.
Fifty-seven of the 547 stars have period changes significant at the 
2-$\sigma$ level or greater.}
\end{figure}

\begin{figure}
\figurenum{7a}
\label{fig7a}
\epsscale{1.00}
\plotone{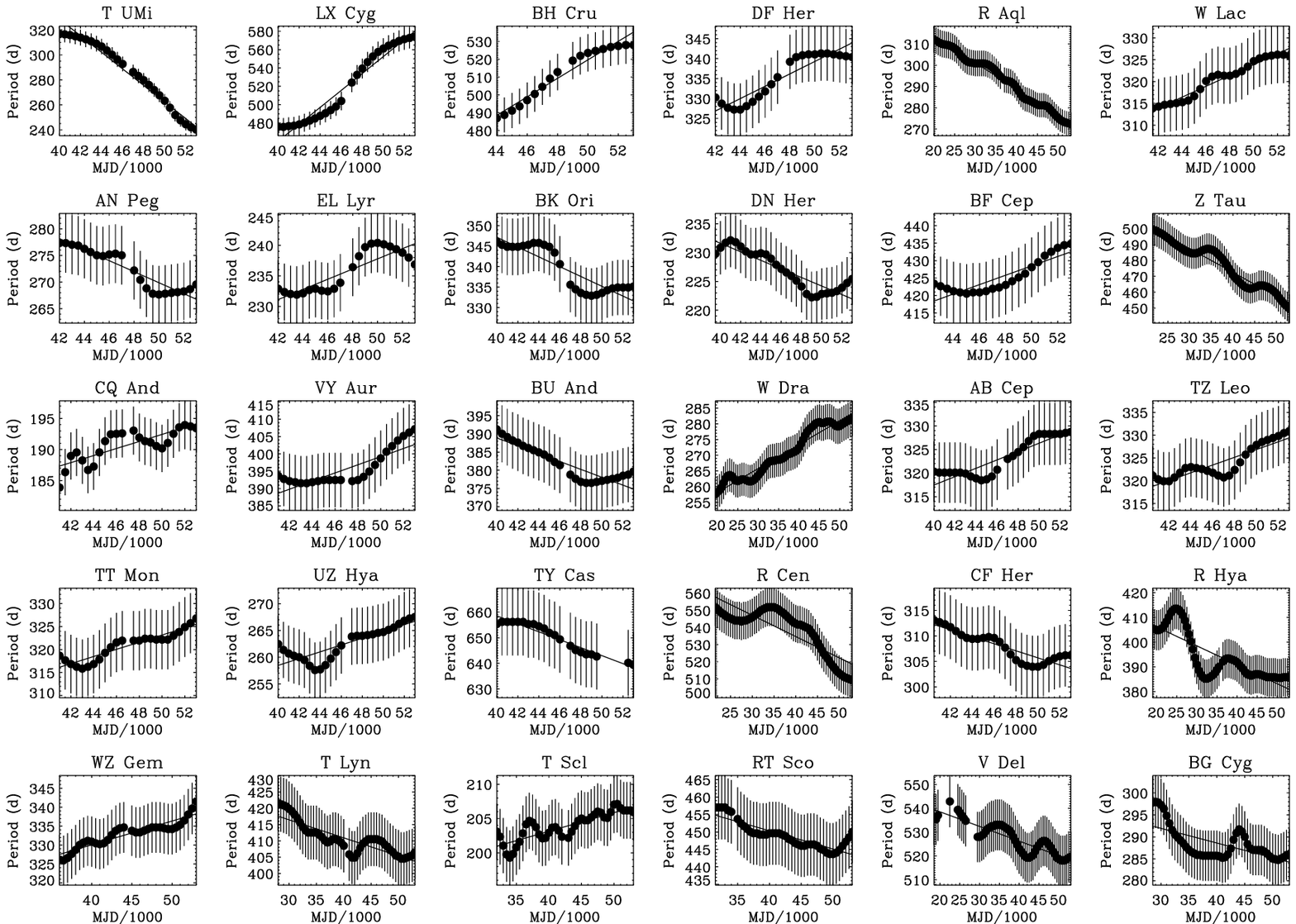}
\caption{Period versus time the 57 stars with non-zero $d\ln{P}/dt$ 
significant at the 2-$\sigma$ level or greater. Solid points are the period at
which the wavelet $Z$-transform is maximal at each value of $t$, separated by
500 days.  Error bars
are 1-$\sigma$, defined by 0.02$P$.  Solid lines are the best linear fit 
through the data points.  Clearly, some stars are not well-fit by a linear
period change, such as RU Tau and S Ori.  Others, like T UMi and LX Cyg
are very well-fit by linear trends.}
\end{figure}

\begin{figure}
\figurenum{7b}
\label{fig7b}
\epsscale{1.00}
\plotone{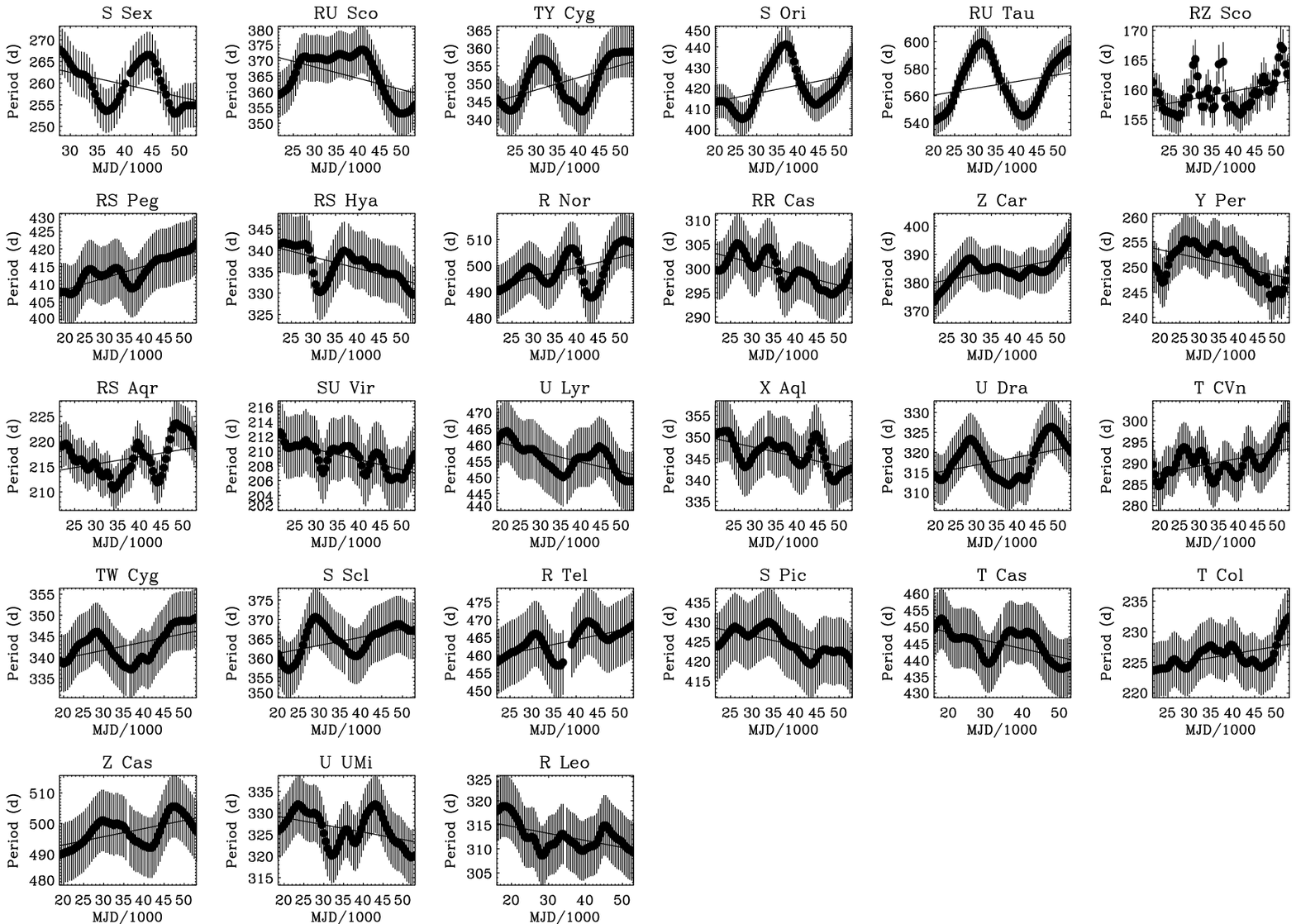}
\caption{Period versus time the 57 stars with non-zero $d\ln{P}/dt$ 
significant at the 2-$\sigma$ level or greater. Solid points are the period at
which the wavelet $Z$-transform is maximal at each value of $t$, separated by
500 days.  Error bars
are 1-$\sigma$, defined by 0.02$P$.  Solid lines are the best linear fit 
through the data points.  Clearly, some stars are not well-fit by a linear
period change, such as RU Tau and S Ori.  Others, like T UMi and LX Cyg
are very well-fit by linear trends.}
\end{figure}

Figure \ref{fig6} shows the measured values of $d\ln{P}/dt$ versus period for 
the 547 stars in the sample.  The first result is that the majority
of stars do not have measurable $d\ln{P}/dt$ significantly different from
zero.  Stars on the AGB should spend most of their time in the interpulse
phase, where $d\ln{P}/dt$ is of order $10^{-5}$.  In fact, we find the average
error ($\sigma_{\rm fit}$) is ten times larger than this, meaning that the
slow increase in period is not generally detectable with our analysis 
technique.  We found 57 stars with $d\ln{P}/dt$ more than 2-$\sigma$ different
from zero, which we list in Table 1.  (The complete Table 1, listing the
average periods and rates of period change for all 547 stars is available 
online.)
Twenty-one of these stars had $d\ln{P}/dt$ significant at the 
3-$\sigma$ level or greater, and eight stars exhibited period changes at the 
6-$\sigma$ level or greater.  Plots of period versus time for these stars are 
shown in Figures \ref{fig7a} and \ref{fig7b}.

The eight stars with the most-significant period changes have all been noted 
previously in the literature.  They are: T UMi, LX Cyg, BH Cru, R Aql, Z Tau,
W Dra, R Cen, and R Hya.  Because the historical literature for these eight
bears summarizing, we discuss and compare our results for each of these stars
with those previously published in the following subsections.  We then discuss
the general results, followed by a short comparison of analysis methods.

\subsection{Rapid period changes}

\subsubsection{T UMi}

\begin{figure}
\figurenum{8}
\label{fig8}
\epsscale{1.00}
\plotone{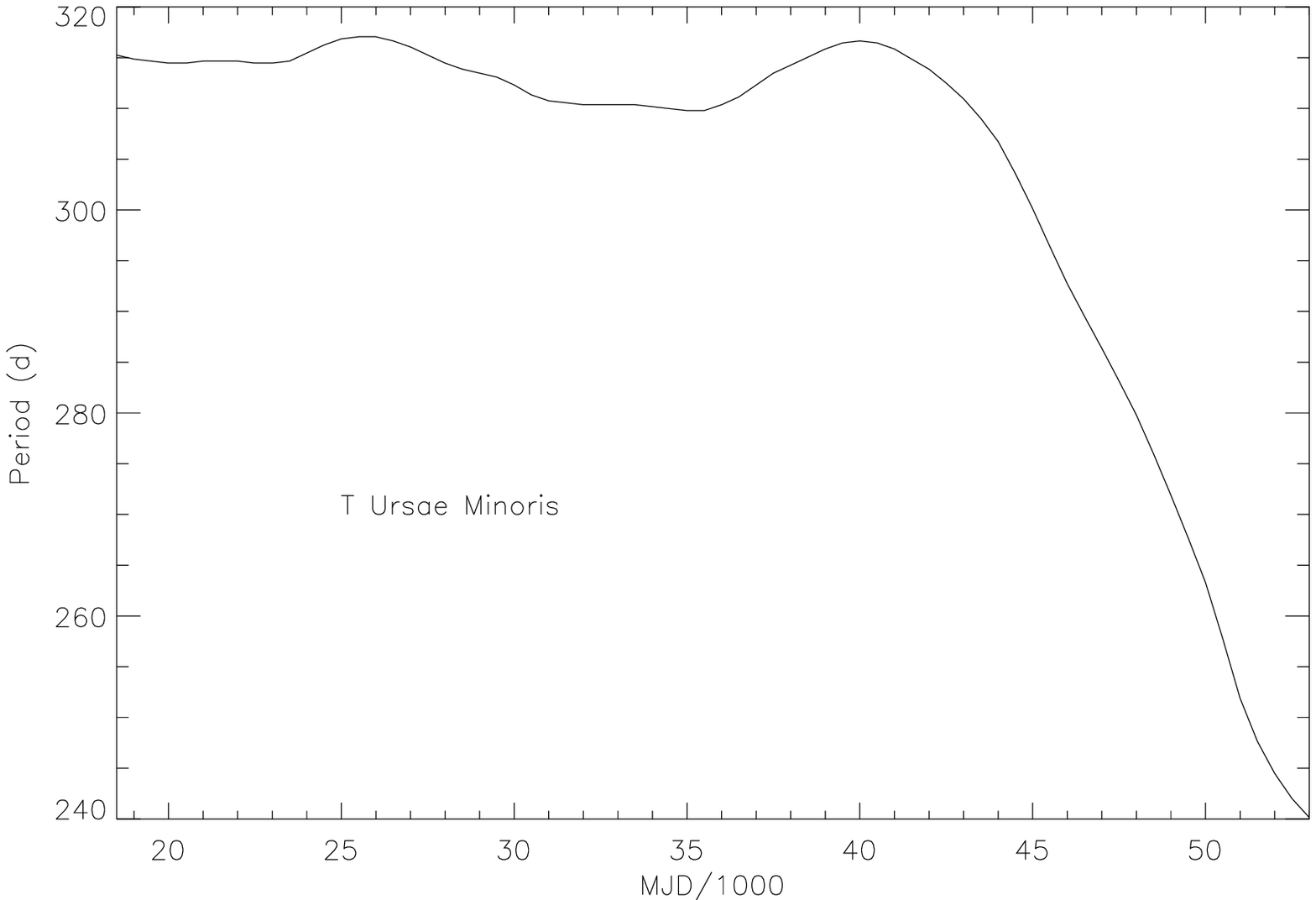}
\caption{Period versus time for T Ursae Minoris, including the entire span of
the data.  The rapid period decline clearly begins midway through the span of
data.  Prior to JD 2440000, the period was essentially constant at 315 days,
but has since fallen to 240 days.  The amplitude (not shown) has declined
along with the period.
}
\end{figure}

T Ursae Minoris is the most dramatic case of a Mira star with a period 
change.  Our analysis of the AAVSO visual data indicates a rate of period 
change of $-8.4 \times 10^{-3}$ y$^{-1}$, the largest known of any Mira 
variable.  The period change began after the star had been monitored and
maintained a constant period for many decades, and size of the period change 
has been very large (nearly 25 percent).  The first note of a modest period 
change (a decline of 10-15 days) appeared in the {\em General Catalogue
of Variable Stars} \citep{gcvs}, which gave a period of 301 days for the epoch
JD2445761.  In 1993, a mention of the behavior of T UMi was made on a variable
star mailing list, and though no archive of that discussion survives, it 
did trigger further investigations.  \citet{pred94} used a period of 280 days
in the calculation of predicted maxima for 1994, indicating that a larger 
period discrepancy was known and measured by January 1994.  The nature of the 
period change was subsequently investigated independently by \citet{gs95} and 
\citet{mf95}, who obtained similar results.  The most recent published 
analysis of T UMi was by \citet{skb03} who interpret the period variations as 
arising from a thermal pulse.  They used a variety of statistical methods to 
measure the period change, but their linear fitting method yielded 
$-3.4 \pm 0.5$ d y$^{-1}$, or $-1.19 \pm 0.18 \times 10^{-2}$ y$^{-1}$ in units
of $d\ln{P}/dt$.  The slight disagreement between their value and ours is 
likely due to their use of a shorter span of data, beginning at JD 
2444000 rather than our JD 2440000.  Our data range includes
the early, shallower portion of the period change onset, and a slightly
smaller average rate is therefore expected.  Otherwise, the values derived are
very similar despite being done with wholly independent data sets,
a good indication that our visual data are consistent with other data
sets, and that our analysis method is sound.  We show our {\em WWZ} results 
for the full span of available AAVSO data for T UMi in Figure \ref{fig8}.

As \citet{skb03} state, the onset and rate of period change are 
entirely consistent with what has been predicted by theoretical evolution 
models of period change in thermal pulses.  However, the modeling of the 
early stages of thermal pulses is difficult.  Most evolution models assume that
the local and global thermal time scales are much shorter than the nuclear time
scale, which is not the case during thermal pulses.  Thus it is difficult
to model the early stages of thermal pulses without resorting to thermonuclear
hydrodynamic models instead.  \citet{wz81} and \citet{vw93} do not explicitly
state the time steps used in their calculations, but we expect their model 
accuracy to be lowest during the earliest phases of the thermal pulse in which
we believe T UMi to be.
The rates of period change for the \citet{vw93} models are on the 
order of $10^{-2}$ y$^{-1}$, consistent with our result.  Because of the 
numerical difficulties in modeling this evolution phase with standard models, 
it is difficult to assess whether the observed rate of period change is in
quantitative agreement with physical expectation.  However, the 
{\it qualitative} behavior of the star is very similar to what is expected 
during this phase, and changes in theoretical models will likely not produce
differences larger than a factor of two.  Therefore, we concur with
the general consensus that the observed period change in T UMi indicates that 
it is a star in the earliest phases of a thermal pulse.

\subsubsection{LX Cyg}

Period change in LX Cyg was first noted by \citet{aa85}, and Mattei 
used periods of 500 days \citep{pred95}, 550 days \citep{pred97}, and 600
days \citep{pred99} to predict dates of maxima for those years.
\citet{b00} derived a period of 568d for the epoch of JD2450377, and the
nature of the period variation was investigated in some detail by
\citet{t03}, who found a smooth variation in period over the entire span of
available observations.  \citet{z04} re-analyzed the available AAVSO data
(also using {\em WWZ}) and confirmed the results of \citet{t03}.  They also 
noted the similarity of LX Cyg to BH Cru (discussed below), and that LX Cyg 
has a transitional spectrum of type SC, indicating carbon enrichment has 
occurred or is ongoing.

Our reanalysis of a slightly longer span of AAVSO visual observations again
clearly shows the period variation, and a linear fit of the period variation
yields $d\ln{P}/dt$ of $+6.5 \times 10^{-3}$ y$^{-1}$.  Published evolution 
models suggest that a period {\it increase} occurs later on in the thermal 
pulse, if indeed LX Cyg is undergoing one.  However, we note that the rate of 
period change in LX Cyg appears to be decelerating.  As Figure \ref{fig7a} 
shows, its
period change has clearly not been linear, and has slowed in recent years.  
Further monitoring of LX Cyg in the coming decades is required, and may help 
to show whether this object is in the later stages of a thermal pulse, or 
whether the period variations are caused by some other phenomenon.

\subsubsection{BH Cru}

Period change in BH Cru was first noted by \citet{b88}.  \citet{w99} later
discussed the spectral behavior and tentatively made the connection between
the changing period found by \citet{b88}, and the {\it spectral} changes
observed over the past 30 years; though BH Cru was originally assigned a
spectral type of SC \citep{cf71}, it now shows carbon bands indicative of
a CS spectral type.  \citet{w99} attributed the period evolution as being
due to a thermal pulse, and argues that the changing spectral type may be
caused by a dredge-up of processed nuclear material in real-time.  However,
\citet{z04} analyzed the visual observations of the {\em Royal Astronomical 
Society of New Zealand} (RASNZ) along with the spectrum of this object, and 
suggested that both the period changes and concurrent spectral changes could 
be due to a global change in effective temperature, rather than a change in 
structure caused by a thermal pulse.  They suggest that feedback between 
opacities and stellar structure could result in structural changes to the 
star, causing a shift in mode period without requiring a thermal pulse.
Thus they raise the possibility that these stars may undergo large period
changes {\it without thermal pulses}.

We analyzed the AAVSO visual observations of BH Cru, and determined the rate of
period change to be $+3.7 \times 10^{-3}$ y$^{-1}$, consistent with the 
\citet{z04} analysis of the RASNZ visual data.  As with LX Cyg, the trend in 
period change is not linear, but has exhibited an almost sinusoidal variation
over the span of available observations.  Such behavior is observed in some 
other Mira stars with long periods, and they have been dubbed ``meandering 
Miras'' by \citet{zb02}.  However, most of the long-period Miras with
unstable periods only vary by a few percent about a mean period, while BH Cru
(and LX Cyg) have changed their periods by over 10 percent over the available
span of observations.  While we agree that both BH Cru and LX Cyg may
simply be manifesting unstable periods, we are not yet prepared to discard
them as thermally-pulsing candidates.  We discuss possible physical mechanisms
for their period variations in Section 5.

\subsubsection{R Aql and R Hya}

R Aql and R Hya have long been known as stars exhibiting large period changes.
(See \citet{mh18} for historical reviews.)  Period changes in these stars were 
detected and discussed by both \citet{p32}
and \citet{sc37}, though neither quantified the rate of change or offered
a physical explanation for them.  Both stars were cited as possible examples 
of thermally-pulsing objects by \citet{wz81}, who obtained reasonable fits 
between the observed period variations and the theoretical period variations 
from evolution model calculations (although they assumed first-overtone 
pulsations).  The rate of period change in R Aql has remained essentially 
unchanged over the available span of observations (see \citet{gh00} for a 
review).  \citet{zbm02} studied both modern and historical data for R Hya 
dating back to 1662.  They found that the historical data indicate
the onset of a near-constant period change beginning around 1770, when the 
period was approximately 495 days.  Their derived rate of period decline is 
$-1.48 \times 10^{-3}$ y$^{-1}$, a factor of two larger than the rate derived 
from the linear fit to the AAVSO data spanning JD 2419389 to 2452894.  This
decline continued until 1950 when the period levelled out at a nearly constant
value, and has remained near 395 days since that time.  Our smaller 
derived value for the period change is a reflection of our shorter span of 
data during which the rate of period change rapidly decelerated.  For this 
star, the longer span of observational data used by \citet{zbm02} is clearly a
better reflection of the long-term evolution of R Hya.  In fact, the behavior 
of R Hya for the latter half of the 20th century resembles the behavior of the
meandering-period stars.

As \citet{wz81} clearly show, the behavior of the periods of both R Aql and R
Hya are reasonably interpreted as the result of radius variations after the
peak luminosity in a thermal pulse.
\citet{wz81} determined the ``time-scale'' for
these stars, equivalent to the inverse of $d\ln{P}/dt$, obtaining values of
950 and 650 years for R Hya and R Aql, respectively.  Our values of
$d\ln{P}/dt$ yield timescales of 1440 and 641 years, respectively.  The value 
for R Aql is in excellent agreement.  The much longer value 
for R Hya is a reflection of the much slower rate of period change 
in the more recent times covered by our data set.  

\subsubsection{W Dra}

W Dra has also long been known to have period variations \citep{mh18}.
The third edition of the {\it GCVS} \citep{k69} noted that the period of
W Dra is variable, and subsequent studies showed that it varied in a 
monotonic fashion.
\citet{wz81} first suggested W Dra as another candidate for a thermal pulse
after analyzing the $O-C$ diagrams of 48 well-observed Mira stars.  Unlike R Hya
and R Aql, the period of W Dra was {\it increasing}, suggesting that it was in
a different stage of the thermal pulse.  \citet{wz81} suggested it is near the
first peak in period following the onset of a thermal pulse, and that the 
period change should cease and/or change sign over the next century.  
\citet{kl95} and \citet{pc99} confirmed that the period change in W Dra was 
real by analyzing times of maximum, and \citet{g00} used the {\em AMPSCAN}
\citep{h91} procedure to measure the period change in this star.  He found
that the period increased from about 255 days in the early part of the 20th
century, to about 280 days in the late 1970's, where it
has remained since.  We obtained similar results, and derived a $d\ln{P}/dt$
of $+1.0 \times 10^{-3}$ y$^{-1}$, though we note we included the most recent
constant-period data in the fit, which underestimates the rate of period change
earlier in the data.  It is not yet clear whether the most recent few decades
of nearly constant period represent a fundamental change in the behavior of
W Dra, or whether it is a short-term excursion within a longer,
more constant trend.  Observations in coming decades will clarify this, and
monitoring of W Dra should continue.

\subsubsection{R Cen}

R Cen has exhibited substantial variations in its light curve shape as well
as its period during the course of its observed history.  R Cen and R Nor are 
the prototypes of the ``double-peaked'' Miras, whose light curves exhibit two 
distinct maxima per cycle.  However, unlike R Nor, the light curve of R Cen 
has undergone significant changes in morphology in recent decades, where the 
double peaks have become far less pronounced at times.  The amplitude
has also decreased with time, and is currently about one third
of the amplitude in the early part of the 20th century.  The true period of
R Cen is also debatable;
\citet{f63} suggested that the difference between the absorption and emission
line velocities in the spectrum of R Cen were consistent with it being a star
with half the GCVS period of 550 days, and this assertion has been adopted
by others since \citep{kgd74,j94}.  However, the characteristics of R Cen are 
very similar to other stars with known or suspected double maxima, and
all of these stars have long periods \citep{l05,tw05}.  \citet{wg01} noted
that the two maxima have different $(B-V)$ colors, indicating that the 
physical causes of each maximum are different, and that the time between
two maxima having the same color is the true period.

The general behavior of the period variation was discussed by \citet{wg01},
and the AAVSO visual data for R Cen between 1950 and 2000 were analyzed in 
detail by \citet{hmf01}.  The latter derived a rate of period change of about 
-1 day per year, yielding $d\ln{P}/dt$ of $-1.9 \times 10^{-3}$ y$^{-1}$.  
Because they only analyzed the most recent data with the strongest period 
decline, their rate is more than twice our value derived using all data from 
1917 to 2002.  Despite the fact that the period was reasonably constant from 
1917 to 1950 and then seems to have begun an accelerated decline like T UMi, 
the behavior of R Cen resembles that of the other long-period stars such as 
LX Cyg and BH Cru.  In fact, both of the latter are considered members of 
the ``double-maxima'' class \citep{gcvs} and so the comparison to these
stars is appropriate.

\subsubsection{Z Tau}

Coherent period change in Z Tau was first noted by \citet{n86}, who analyzed 
AAVSO $O-C$ data covering the period of 1915 to
1983.  \citet{pa99} did not give a specific rate of period change but note 
that the rate of decline was larger than $10^{-3}$ days day$^{-1}$;
the corresponding value of $d\ln{P}/dt$ would be $7.7 \times 10^{-4}$ y$^{-1}$.
Using AAVSO visual data from 1917 to 2003, we derive an average value of
$-1.15 \times 10^{-3}$ y$^{-1}$, fully consistent with the lower limit of
\citet{pa99}.  This rate of period change appears to be relatively constant 
through the span of available data.  It is also consistent with theoretical
rates derived from models of the later stages of thermal pulses.

\subsection{Slower period changes}

Forty-nine more of 547 stars showed period variations whose linear fits were
statistically different from zero at the 2-$\sigma$ level, with 13 of these 
being significant at the 3- to 6- $\sigma$ level.  For many of these stars, the
period variations are not linear in nature, and the use of a linear fit is 
clearly not representative of their behavior; its utility is in 
assessing the {\it general trend} of period change throughout the data set.  
(A few stars, notably DU Cyg and S Tri, exhibited modest period changes 
throughout the span of observations, but their {\it net} period change was 
small enough that they did not meet our selection criteria.)  Among the stars 
that {\it do not} exhibit clear linear trends in period are several whose 
periods appear to vary in a seemingly sinusoidal fashion.  The best examples 
of this behavior are RU Tau and S Ori, with S Sex, TY Cyg, and U Dra being 
slightly weaker examples.  \citet{zb02} described such objects as ``meandering
Miras,'' suggesting that as many as 10-15\% of all stars of long period may 
exhibit such behavior.  Such behavior, on timescales of {\it decades} rather 
than centuries, is not well-modeled by evolutionary processes, and may be a 
manifestation of some other type of instability.  We discuss this further in 
section 5.

At least one star in our sample, Y Per, underwent a drastic change in pulsation
behavior, switching from Mira-like variations to semi-regular variations in 
{\it less than two pulsation cycles}.  This change was also coincident with a 
modest drop in period.  This behavior was first noted by \citet{k00}, who 
suggested that vigorous convection may be at work.  \citet{c91} noted that
the semi-regular stars RV And, S Aql, and U Boo exhibited significant 
variations in pulsation period and amplitude, and suggested that 
mode-switching could explain the behavior.  Other Miras have changed 
from regular pulsators to semi-regular or irregular ones in the past, notably
W Tau and RT Hya \citep{mmw90}, both of which have similar periods to Y Per.  
We note these two objects appear to be re-establishing more regular pulsations,
but more observations are required to confirm this.  The transition from Mira 
to semi-regular may be transient in nature, as may changes in period.  Objects
like Y Per, W Tau, and RT Hya should be continually monitored in the future.

A large fraction of the stars showing ``variations'' are marginal detections,
between 2- and 3-$\sigma$.  
While such objects
meet the statistical requirement for mention in our results, the period 
variations appear to be erratic at best, and do not represent a coherent
change in period; CQ And and RZ Sco are particularly notable examples of this.
We suggest that for many of these stars, the
measured ``period changes'' are simply manifestations of the random period
changes known to occur in Mira stars.  Miras can have substantial
jitter in period from cycle to cycle, and it is reasonable to assume that
in any large sample of stars, some objects may exhibit period variations
large enough to be flagged by our criteria.

\subsection{The utility of wavelet analysis for deriving rates of period
change}

Our work
has highlighted the many different methods for searching for period changes in
data.  The classical method for performing such studies was the use of
maxima and minima data to create $O-C$ diagrams.  Linear trends in
period appear as parabolic curves in $O-C$, making the detection of period
changes reasonably straightforward.  This method was employed in the earliest
work of Eddington and Plakidis in the 1920's and 1930's, and continues to
be employed today.  See the series of papers by \citet{kl93} and Percy et 
al.~\citep{pa99,pc99} for more details.  The major limitation of such studies 
is that 
they require the measurement of times of maxima and minima -- in effect a 
``pre-processing'' of the data.  More computationally-intensive
methods such as wavelet, or other ``time-frequency'' analyses allow us to use
the data directly.  Such methods are not perfect; variations in light curve
shape, long known to affect the Mira variables, can have a substantial effect
on the Fourier spectra of these objects, and the additional computational time
required for such methods has only become ``trivial'' relatively recently.
As the extensive work by both Koen \& Lombard and Percy et al.~shows, our 
analysis method and others similar to it should be considered 
{\it complimentary} to $O-C$ and time-of-maximum analyses, rather than be 
viewed as replacements.

One clear example of this is in the detection of period changes with
very long data sets composed only of times of maximum, when actual 
observations are not available.  \citet{sbk99} used
$O-C$ analysis to detect a small but significant period change in $\chi$ Cyg
dating to its discovery 1686.  The data from the 20th century has
period changes fully consistent with the random variations seen in some Mira
variables, and since $\chi$ Cyg is a long-period star, such period variations
are not surprising.  Our data and the \citet{sbk99} data set produce similar
results in this regard.  However, when the time-of-maximum data were analyzed,
longer term trends began to emerge.  The full 320-year $O-C$ diagram indeed
revealed a parabolic curve, and \citet{sbk99} determined that the rate of 
period change is consistent with the rate expected in the late stages of a 
thermal pulse, after the hydrogen-burning shell has fully re-established 
itself.  \citet{pa99} also suggest that even recent time-of-maximum data
can be used to detect the long-term period evolution in Mira stars, though
as they state, the statistical significance of their result is less than
2-$\sigma$.  Unfortunately, historical data are unlikely to exist for a
significant number of Mira variables, since only a handful were discovered
prior to the late 19th century.

\section{Discussion}

The incidence of statistically significant, long-term period variations in our
sample is on the order of ten percent, and we see large changes in about one
percent of our sample.   The latter fraction is reasonable given the expected
durations of the rapid shell-flashing phase and the slower inter-pulse 
phase.  Because our sample is a 
heterogeneous mixture of stars with different ages and masses, it is not 
possible to make an exact prediction of the numbers of stars we expect to 
undergo such changes.  The physical mechanism for period changes in Mira 
variables has not been proven, though we expect thermal pulses to occur on the
AGB, and we expect that some stars must be Mira variables when these pulses 
occur.  Therefore, we expect to observe changes in pulsation behavior in at 
least some Mira variables, given a large enough sample size.  Below, we discuss
potential
evolutionary and non-evolutionary mechanisms for the observed period changes in
our sample.

\subsection{Evolutionary period changes}

From theoretical models of thermal pulses, we expect to see large and
relatively monotonic variation in period on timescales of centuries.  This
is indeed what we see for eight stars of the 547, although even for some
of those, there are secondary period variations apparent on the
period vs time curves.  Thus, the interpretation of these secular changes
as resulting from shell flashing appears both natural and appropriate.
However, the fact that secondary variations {\it do appear}, and that other
stars {\it only} show these secondary, meandering variations leaves room for
either a different or modified explanation for large, secular period 
variations.

\begin{figure}
\figurenum{9}
\label{fig9}
\epsscale{1.00}
\plotone{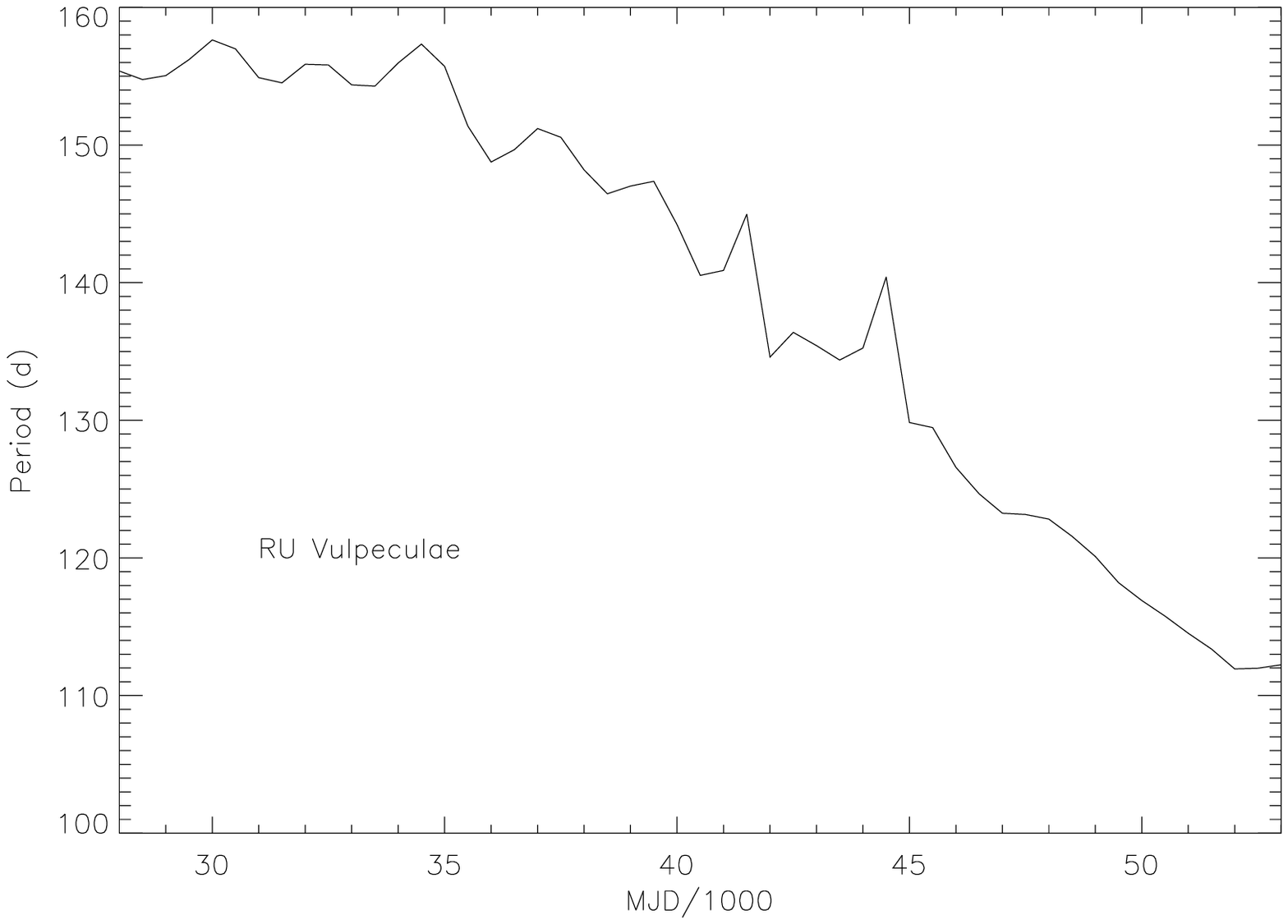}
\caption{Period versus time for RU Vulpeculae, a semi-regular variable with
an apparent period change.  The period has declined from approximately 155 days
to about 110 days at present, a decline of nearly 30 percent in 65 years.
The amplitude of pulsation has dramatically declined as well, and the star
may cease pulsating altogether in the coming decades.
}
\end{figure}

Even if the thermal pulse explanation is correct, it is then difficult to 
estimate the number of Miras we expect to exhibit strong period changes in a 
given population.  The Miras in our sample are a mixture
of spectral types and must also have a range of masses based upon the
range of periods observed.  If these stars only spend part of their AGB
lifetimes within the Mira instability strip, then we may see period changes
in other types of AGB variables, such as semi-regulars and obscured carbon
and OH-IR stars, all of which have similar luminosities as the Miras.  We
are currently analyzing data for the semi-regular variables found in the
AAVSO ID to search for period changes, but note the existence of at least 
one well-known semi-regular variable with a period change, namely RU Vul,
which we show in Figure \ref{fig9}.  The large period change was first noted
by \citet{zb02}, and though the star was noted as having a double period
by \citet{k99}, both its period change and its declining pulsation amplitude 
are remarkably similar to those of T UMi.  Continued monitoring of this star in 
the coming years is strongly encouraged.

According to theoretical models of \citet{vw93},
stars with solar metallicity and masses larger than 1 M$_{\odot}$ 
spend between 70,000 and 100,000 years between thermal pulses.  The pulses
themselves cause large changes in period for perhaps 2,000 years, and thus
the ratio of pulse to interpulse lifetimes is between 2 and 3 percent.  If
the Miras are evenly distributed throughout the thermal-pulsing AGB, then
we might expect large, secular period changes in the same percentage of
Miras.  Eight of 547 stars represents $1.6\%$ of the sample, in
reasonable agreement with this expectation.

We note that when AGB stars undergo a thermal pulse, they may become or cease
to be Mira stars for a period of time.  The models only lie within the
Mira instability region for part of the thermally-pulsing and interpulse
phases of AGB evolution.  During the pulse, transient oscillations may
develop in the atmosphere, turning the stars into semi-regular
variables, or quenching the pulsations altogether.  The thermal pulse may also
be accompanied by mass-loss events, turning the stars into extreme carbon or 
OH-IR stars, and rendering them optically faint or invisible.  In the former 
case, period changes may indeed be visible among the semi-regular variables, 
and we are currently analyzing the semi-regular variables found in the AAVSO 
ID for a forthcoming paper.

\subsection{Non-evolutionary period changes}

Several stars in our sample exhibited non-monotonic period variations with
time-scales of a few decades, either about the mean period, or superimposed
upon a larger trend.  The most prominent examples of this are RU Tau and S
Ori, whose period variations appear to be nearly sinusoidal.  Although we 
cannot definitively claim the period variations are truly oscillatory in
nature (because the timescales are of the same order as the span of the data),
the ``periods'' of these variations are on the order of 10 to 80 years.
We do not have an immediate explanation for why such variations exist, but we
speculate that they may be related to thermal pulses.  While the 
decades-long
timescales of these variations are far shorter than the those predicted 
for pulse-induced global changes, they may represent {\it thermal oscillations}
in the envelope with characteristic Kelvin-Helmholtz timescales, $\tau_{KH}$,
on the order of a few decades.  The Kelvin-Helmholtz timescale can be derived 
if one can reliably estimate the radial order of the pulsation mode, the 
luminosity, and the mass of the Mira star.  The precise value of $\tau_{KH}$ 
for individual stars must be derived using known values of stellar parameters,
and for most objects in our sample, these values are not known.  Studies of
mass loss and evolution (see \citet{bw91} for example) allow us to 
{\it estimate} the mass based upon the period, and for nearly all reasonable 
values of these quantities, $\tau_{KH}$ is on the order of a few decades.

We can estimate the Kelvin-Helmholtz timescale, $\tau_{KH}$, as

\begin{equation}
\tau_{KH} \sim {{GM^{2}}\over{RL}} \sim 3.1 \times 10^{7} 
{{(M/M_{\odot})^{2}}\over{(R/R_{\odot})(L/L_{\odot})}}  {\rm y.}
\end{equation}

\noindent
\citet{oc86} calculated many AGB models in the range of 0.8 to 2.0 
M$_{\odot}$, and obtained Kelvin-Helmholtz timescales between 6 and 200
years.  A Mira star with $M=1.0M_{\odot}$, $L=5000L_{\odot}$, and 
$R=250R_{\odot}$ would have $\tau_{KH}$ of $\sim 25$ years, in good agreement
with the observed timescales.  Thus we suggest that the shorter-timescale 
period variations may be thermal relaxation oscillations in the stellar 
envelope, as it responds to the global changes caused by a thermal pulse.
A more comprehensive study of these variations will appear in a future paper.

We also note two additional non-evolutionary explanations for such behavior.
One is that Mira variable pulsations are intrinsically non-linear in nature,
and that they exhibit low-dimensional chaotic behavior.  \citet{ks02} argue 
that amplitude (not period) variations in the light curve of R Cygni are caused
by low-dimensional chaos, perhaps through the non-linear interaction of two or
more pulsation modes.  Such a mechanism was suggested by \citet{b96}
to describe the irregular behavior of RV Tauri-type stars, and a similar 
(albeit less irregular) mechanism may be at work in the Mira stars.

A second explanation for large period changes was put forth by \citet{yt96}, 
who suggested that Mira pulsations may be strong enough to modify the interior 
structure as the star pulsates.  They suggest that the pulsations modify the 
entropy structure of the star over time, causing both mode-switching and a 
readjustment of the equilibrium structure of the star.  Their theory was tested
with non-linear, hydrodynamic models in which a fundamental-mode Mira model
changed to first-overtone pulsation.  The ``new'' model period was 
significantly different from the initial equilibrium first-overtone period as
a result of structural changes wrought by the strong pulsations.  More modeling
and investigations along both of these lines would be useful, and may explain 
several other aspects of Mira variability such as the small, cycle-to-cycle 
variations in period and amplitude.

In closing, we also note that some of the stars with ``long-term'' period 
changes may in fact be exhibiting the same behavior as stars showing these 
shorter-term variations, but on a longer time-scale.  In particular, LX 
Cyg, BH Cru, and R Hya seem to be nearing plateaus or troughs in their 
periods, and may indeed begin to change again in the future. Such stars bear 
close watching in the coming decades, both in regards to their periods and to 
their spectra.  It is entirely possible that LX Cyg, BH Cru, and R Hya are
undergoing similar behavior to RU Tau, and that we simply do not have a span of
observations long enough to detect this.  Continued monitoring of these stars
is necessary and encouraged, as the coming decades may clarify their behavior.

\section{Conclusions}

We have performed a wavelet analysis of 547 well-observed Mira variables from
the AAVSO International Database.  We are unable to detect long-term variations
in period in the majority of the stars having $d\ln{P}/dt$ below our detection
threshold.  About 10\% of the sample shows 
period variations on timescales of decades at the 2-$\sigma$ level or greater
and only eight of the 547 (1.6\%) have highly significant ($>6$-$\sigma$)
monotonic period changes.  This latter percentage is consistent with the 
fraction of stars we expect to be in the early 
post-thermal pulse phase at any given moment, during which period change
is predicted to be greatest.  One star, T Ursae Minoris, showed a sudden
onset of large period decrease which is consistent with the onset of a
thermal pulse during the span of observations.  In several stars, we detect 
significant but non-monotonic ``meandering'' changes.  The physical mechanism
for such variations in period is not clear, but the timescale for these 
variations is similar to the Kelvin-Helmholtz cooling timescale for the 
envelope.

Continued monitoring of Mira stars and other long-period variables is clearly
warranted, and will almost certainly continue given the level of interest in
these stars among both the amateur and professional astronomical communities.
The behavior of stars such as T Ursae Minoris suggests that even 
``well-behaved'' stars with nearly constant periods are deserving of
continued monitoring given the possibility of detecting the sudden onset of
period change.  However, the lack of thus-far undetected large-scale period
changes among well-observed Miras suggests that few additional 
thermally-pulsing candidates will be found among Mira stars in the solar
neighborhood, at least in the short-term.

In addition to continued monitoring of known Mira stars, we encourage the 
continued long-term monitoring of the
large numbers of Mira and Mira-like variables detected in the Galactic center
and in other galaxies, since they may produce additional candidates for
thermal pulses once sufficient data has been collected to measure period
changes.  Such monitoring efforts will require decades-long commitments
of telescope time, but this may be easily achievable with many of the current
and planned all-sky monitoring projects, as well as with the continuing efforts
of volunteer and amateur observers.

\acknowledgements
Dr. Janet Mattei passed away shortly before the completion of this work, and her
coauthors respectfully dedicate it to her memory.  We are indebted to the many
thousands of variable star observers worldwide whose work and dedication over 
the past century made this research possible.  We wish to thank Rebecca Turner
and Elizabeth Waagen of the AAVSO for assistance with the AAVSO Mira data, and 
Steven Kawaler, Kevin Marvel, and Arne Henden for helpful comments.  This
research was partially supported by a grant from NASA administered by the
American Astronomical Society.

\newpage

\begin{deluxetable}{rrrrrr}
\tablenum{1}
\tablecolumns{6}
\tablewidth{0pc}
\tablecaption{Measured periods and rates of period change for 57 Mira stars
with $d\ln{P}/dt$ greater than 2-$\sigma$ above the measurement error.}
\tablehead{\colhead{Name} & \colhead{$\bar{P}$} & \colhead{$d\ln{P}/dt$} &
\colhead{N-$\sigma$} & \colhead{$d\ln{P}$} &
\colhead{$(d\ln{P}/dt)^{-1}$}\\
\colhead{} & \colhead{(d)} & \colhead{($10^{-3}$ y$^{-1}$)} & \colhead{} &
\colhead{} & \colhead{(y)}
}
\startdata
    T UMi & $285.49 \pm  1.10$ & $-8.47 \pm 0.35$ & 23.90 & 0.27 &   120 \\
   LX Cyg & $520.15 \pm  2.01$ & $ 6.47 \pm 0.36$ & 17.86 & 0.19 &   150 \\
    R Aql & $293.00 \pm  0.72$ & $-1.56 \pm 0.09$ & 17.02 & 0.14 &   640 \\
    Z Tau & $476.63 \pm  1.19$ & $-1.15 \pm 0.10$ & 11.66 & 0.10 &   870 \\
    W Dra & $270.56 \pm  0.66$ & $ 1.03 \pm 0.09$ & 11.42 & 0.09 &   970 \\
    R Cen & $538.14 \pm  1.35$ & $-0.84 \pm 0.10$ &  8.63 & 0.08 &  1190 \\
    R Hya & $393.89 \pm  0.95$ & $-0.71 \pm 0.09$ &  8.01 & 0.07 &  1410 \\
   BH Cru & $511.87 \pm  2.35$ & $ 3.70 \pm 0.61$ &  6.09 & 0.08 &   270 \\
    V Del & $528.85 \pm  1.46$ & $-0.43 \pm 0.10$ &  4.08 & 0.05 &  2350 \\
    S Ori & $419.99 \pm  1.03$ & $ 0.35 \pm 0.09$ &  3.81 & 0.09 &  2870 \\
   TY Cyg & $350.60 \pm  0.86$ & $ 0.36 \pm 0.09$ &  3.79 & 0.05 &  2790 \\
   RU Sco & $365.28 \pm  0.91$ & $-0.36 \pm 0.10$ &  3.70 & 0.06 &  2770 \\
   DF Her & $335.40 \pm  1.40$ & $ 1.69 \pm 0.46$ &  3.69 & 0.04 &   590 \\
   BK Ori & $339.38 \pm  1.31$ & $-1.29 \pm 0.36$ &  3.56 & 0.04 &   780 \\
    T Lyn & $410.63 \pm  1.15$ & $-0.49 \pm 0.14$ &  3.53 & 0.04 &  2030 \\
   RU Tau & $568.74 \pm  1.39$ & $ 0.32 \pm 0.09$ &  3.52 & 0.10 &  3090 \\
   DN Her & $226.97 \pm  0.86$ & $-1.19 \pm 0.34$ &  3.49 & 0.04 &   840 \\
   RS Peg & $413.72 \pm  0.99$ & $ 0.29 \pm 0.09$ &  3.40 & 0.03 &  3400 \\
    W Lac & $320.47 \pm  1.31$ & $ 1.41 \pm 0.43$ &  3.28 & 0.04 &   710 \\
   RZ Sco & $159.32 \pm  0.40$ & $ 0.31 \pm 0.10$ &  3.24 & 0.08 &  3190 \\
   AN Peg & $272.46 \pm  1.14$ & $-1.39 \pm 0.46$ &  3.02 & 0.04 &   720 \\
    Y Per & $250.77 \pm  0.60$ & $-0.25 \pm 0.08$ &  2.97 & 0.05 &  4000 \\
   BG Cyg & $288.35 \pm  0.82$ & $-0.42 \pm 0.14$ &  2.94 & 0.05 &  2360 \\
    R Nor & $498.20 \pm  1.25$ & $ 0.29 \pm 0.10$ &  2.89 & 0.04 &  3490 \\
   RR Cas & $299.48 \pm  0.75$ & $-0.28 \pm 0.10$ &  2.88 & 0.04 &  3520 \\
   RS Hya & $336.45 \pm  0.85$ & $-0.29 \pm 0.10$ &  2.86 & 0.04 &  3460 \\
   BU And & $381.86 \pm  1.47$ & $-1.03 \pm 0.36$ &  2.85 & 0.04 &   970 \\
    S Sex & $259.52 \pm  0.73$ & $-0.40 \pm 0.14$ &  2.85 & 0.06 &  2520 \\
   WZ Gem & $332.88 \pm  1.13$ & $ 0.69 \pm 0.24$ &  2.83 & 0.05 &  1450 \\
   EL Lyr & $235.69 \pm  0.98$ & $ 1.29 \pm 0.46$ &  2.81 & 0.04 &   780 \\
   AB Cep & $323.24 \pm  1.24$ & $ 1.00 \pm 0.36$ &  2.77 & 0.03 &  1000 \\
    Z Car & $384.51 \pm  0.97$ & $ 0.28 \pm 0.10$ &  2.73 & 0.06 &  3630 \\
    T Scl & $203.82 \pm  0.62$ & $ 0.49 \pm 0.18$ &  2.72 & 0.04 &  2050 \\
    U Lyr & $455.88 \pm  1.10$ & $-0.24 \pm 0.09$ &  2.70 & 0.03 &  4200 \\
   CQ And & $190.73 \pm  0.76$ & $ 1.09 \pm 0.40$ &  2.69 & 0.05 &   920 \\
   VY Aur & $395.69 \pm  1.58$ & $ 1.08 \pm 0.41$ &  2.66 & 0.04 &   920 \\
    T Cas & $444.74 \pm  1.03$ & $-0.20 \pm 0.08$ &  2.61 & 0.03 &  4930 \\
    T CVn & $290.32 \pm  0.69$ & $ 0.22 \pm 0.08$ &  2.61 & 0.05 &  4520 \\
   TZ Leo & $324.09 \pm  1.27$ & $ 0.96 \pm 0.38$ &  2.50 & 0.03 &  1040 \\
   SU Vir & $209.26 \pm  0.52$ & $-0.24 \pm 0.10$ &  2.48 & 0.03 &  4180 \\
    U Dra & $318.09 \pm  0.77$ & $ 0.22 \pm 0.09$ &  2.46 & 0.05 &  4490 \\
   RS Aqr & $216.59 \pm  0.54$ & $ 0.25 \pm 0.10$ &  2.46 & 0.06 &  4080 \\
   UZ Hya & $262.60 \pm  1.01$ & $ 0.88 \pm 0.36$ &  2.44 & 0.04 &  1130 \\
    X Aql & $346.07 \pm  0.86$ & $-0.23 \pm 0.09$ &  2.41 & 0.03 &  4390 \\
   RT Sco & $449.17 \pm  1.37$ & $-0.43 \pm 0.18$ &  2.40 & 0.03 &  2350 \\
   TW Cyg & $342.86 \pm  0.83$ & $ 0.21 \pm 0.09$ &  2.37 & 0.04 &  4760 \\
   BF Cep & $425.46 \pm  1.81$ & $ 1.15 \pm 0.49$ &  2.34 & 0.03 &   870 \\
    S Scl & $364.65 \pm  0.89$ & $ 0.21 \pm 0.09$ &  2.27 & 0.04 &  4780 \\
    Z Cas & $497.31 \pm  1.20$ & $ 0.20 \pm 0.09$ &  2.22 & 0.03 &  5120 \\
    U UMi & $326.27 \pm  0.78$ & $-0.19 \pm 0.09$ &  2.21 & 0.04 &  5250 \\
    R Leo & $312.56 \pm  0.72$ & $-0.17 \pm 0.08$ &  2.20 & 0.03 &  5820 \\
   TT Mon & $320.83 \pm  1.28$ & $ 0.89 \pm 0.41$ &  2.19 & 0.03 &  1130 \\
   CF Her & $307.93 \pm  1.21$ & $-0.82 \pm 0.38$ &  2.13 & 0.03 &  1230 \\
    S Pic & $424.66 \pm  1.06$ & $-0.21 \pm 0.10$ &  2.08 & 0.03 &  4860 \\
    R Tel & $463.65 \pm  1.19$ & $ 0.21 \pm 0.10$ &  2.04 & 0.03 &  4860 \\
   TY Cas & $648.42 \pm  2.87$ & $-0.87 \pm 0.43$ &  2.01 & 0.03 &  1140 \\
    T Col & $226.05 \pm  0.57$ & $ 0.20 \pm 0.10$ &  2.00 & 0.04 &  5050
\enddata
\tablecomments{$d\ln{P}/dt$ is defined as $dP/dt/\bar{P}$, where $dP/dt$ and
$\bar{P}$ are obtained from the linear fit.  $d\ln{P}$ is defined as
$(P_{\rm max} - P_{\rm min})/\bar{P}$.  The error in $d\ln{P}$ is
approximately constant at $\sim 0.04$.}
\end{deluxetable}
\end{document}